\documentclass[fleqn,usenatbib]{mnras}

\usepackage{newtxtext,newtxmath}
\usepackage[T1]{fontenc}
\DeclareRobustCommand{\VAN}[3]{#2}
\let\VANthebibliography\thebibliography
\def\thebibliography{\DeclareRobustCommand{\VAN}[3]{##3}\VANthebibliography}
\usepackage{graphicx}
\usepackage{amsmath}
\usepackage{soul}

\title[Enhanced Destruction by Major Mergers]{Enhanced Destruction of Cluster Satellites by Major Mergers}

\author[K. Dong et al.]{
Kyung Lin Dong$^{1}$, Rory Smith$^{2}$, Jihye Shin$^{3}$, and Reynier Peletier$^{4}$
\\
$^{1}$ Institute of Physics, University of Amsterdam, Science Park 904, NL-1098 XH Amsterdam, the Netherlands\\
$^{2}$ Departamento de F\'{i}sica, Universidad T\'{e}cnica Federico Santa Mar\'{i}a, Vicu\~{n}a Mackenna 3939, San Joaqu\'{i}n, Santiago, Chile\\
$^{3}$ Korea Astronomy and Space Science Institute (KASI), 776 Daedeokdae-ro, Yuseong-gu, Daejeon 34055, Republic of Korea\\
$^{4}$ Kapteyn Astronomical Institute, University of Groningen, Landleven 12, NL-9747 AD Groningen, the Netherlands
}

\date{Accepted for publication in MNRAS 2023 November 29}
\pubyear{2023}

\begin{document}
\label{firstpage}
\pagerange{\pageref{firstpage}--\pageref{lastpage}}
\maketitle

\begin{abstract}
Using a set of clusters in dark matter only cosmological simulations, we study the consequences of merging of clusters and groups of galaxies (with mass ratio larger than 5:1) to investigate the tidal impact of mergers on the satellite halos. We compare our results to a control sample of clusters that have had no major mergers over the same time period. Clusters that undergo major mergers are found to have a significant enhancement in destruction of their subhalos of $\sim$10-30\%, depending on how major the merger is. Those with mass ratios less than 7:1 showed no significant enhancement. The number of destroyed subhalos are measured for the cluster members that were inside the virial radius of clusters \textit{before} the merger begins. This means preprocessed galaxies brought in by the merger are deliberately excluded, allowing us to clearly see the enhanced destruction purely as a result of the distorted and disturbed tidal field of the cluster during the merger. We also consider secondary parameters affecting the destruction of those satellites but find that the major mergers are the dominant factor. These results highlight how major mergers can significantly impact the cluster population, with likely consequences for the formation of intracluster light, and enhancement of tidal features in the remaining satellites.
\end{abstract}

\begin{keywords}
galaxies: clusters: general -- galaxies: evolution -- galaxies: groups: general
\end{keywords}

\begin{figure*}
    \includegraphics[width=\hsize]{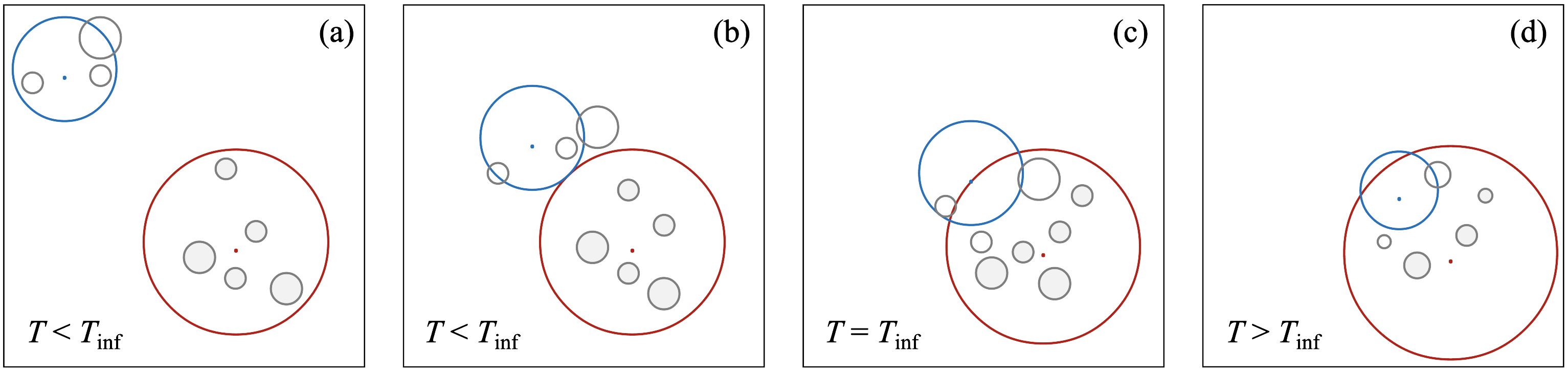}
    \caption{Time series of a merging event between a galaxy cluster (red circle) and a group of galaxies (blue circle). Grey dots represent the member galaxies of the cluster, if shaded, and the group, if not filled. The size of objects is arbitrary. Figure 1(a) indicates the time prior to the merger, 1(b) $T_\mathrm{overlap}$ when the boundaries of the cluster and group halos start to overlap, 1(c) $T_\mathrm{inf}$ when the centre of the group host halo crosses the virial raidus of the cluster and the merger actually happens, and 1(d) the time after the two halos experience tidal interactions.}
    \label{merger_cartoon}
\end{figure*}

\section{Introduction}
\label{sec:intro}

Clusters of galaxies are some of the most dense and massive virialised structures, that are actively growing in the Universe today. They have typical sizes of several Mpc and typical masses ranging from $\sim10^{14}$ to $10^{15}\; \mathrm{M}_\odot$. Galaxy clusters are filled with hundreds to thousands of galaxies, all of which are bathed in the intracluster medium (ICM), a hot ionised gas of temperature $\sim 10^8$ K \citep{Sarazin1986} which emits in the X-rays. 

Galaxy clusters contain a Brightest Cluster Galaxy (BCG) which is typically the central galaxy of the cluster, and thus often grows by consuming other galaxies during major mergers \citep{Merritt1985,Katayama2003}. Indeed, simulations have shown that cluster mergers can greatly enhance the stellar mass of the BCGs and even induce a small amount of new star formation \citep{Contreras-Santos2022}. The BCG is usually surrounded by an envelope of diffuse light known as the Intra-Cluster Light (ICL) \citep{Rudick2006,Burke2013,Contini2013,Montes2019,Contini2021}. The ICL is the diffuse light that arises from stars that were stripped out of member galaxies. Its existence was predicted by \cite{Zwicky1937} and later confirmed by observations of the Coma Cluster \citep{Zwicky1951}. One possibility for the origin of the ICL is due to the tidal interactions between galaxies and clusters \citep{Gregg1998} even if they may not be the dominant source \citep{DeMaio2015}. These two features, the BCG and the ICL, can provide important clues about the overall growth history of a galaxy cluster.

Being located at the nodes of the cosmic web, galaxy clusters are typically attached to cosmic filaments. These represent highways along which clusters are fed fresh mass \citep{Rost2020} and, as a result, tend to grow preferentially from the directions of the filaments \citep{Kuchner2021,Smith2023}. Studies have shown that the connection of a cluster to its filaments can affect the cluster properties including cluster shape and dynamics \citep{Codis2018,Song2018,Welker2018,kraljic20,Gouin2021,Smith2023}. Furthermore, the properties of galaxies that are fed in from filaments are also different than non-filament galaxies including colours, ages, metallicities, $\alpha$-abundances and star formation rates \citep{Bahe2013, kraljic20, Winkel2021}. This may be, in part, due to environmental effects acting within the filaments such as ram pressure stripping \citep{Benítez-Llambay2013, Thompson2023}. Additionally, \cite{Song2021} found that the high vorticity in the region surrounding filaments can disrupt the gas inflow onto galaxies that pass through it.

Galaxy groups tend to be preferentially located within filaments. Thus, filaments are the likely source by which groups of galaxies are fed into clusters \citep{Bekki1999,Kuchner2021}, resulting in major mergers if the masses of the groups are comparable to those of the central clusters. While merging, the satellites that were originally virialized in their clusters and groups can be destroyed or tidally perturbed and thrown to the core or outskirts of the clusters \citep{Choque2019}. These major mergers can heavily disturb a cluster's dynamics and transform their overall morphologies. The shock to a cluster's ICM can be directly observed in the form of radio relics. The shocks and turbulence generated by mergers accelerate relativistic particles in the ICM which, in the presence of magnetic field in galaxy clusters, result in synchrotron radiation \citep{vanWeeren2011,Hoeft2012} which appears as a radio relic \citep{Hoeft2004,Ferrari2008,Cassano2010,vanWeeren2011b,Feretti2012,Lindner2014}. The orientation of the radio relic can provide useful information on the direction of the merger, and the mass ratio of the merger. The distance from the cluster centre of the radio relic provides information on the time since the merger. If the merger is sufficiently major, the ICM may separate from the mass distribution of the cluster due to its collisional nature. A very well known example of this can be seen in the Bullet Cluster \citep{Mehlert2001,Markevitch2002,Markevitch2004,Clowe2006}. Large samples of merging clusters have been gathered and simulations are used to try to best constrain the merging scenario \citep{Chatzikos2010,zuhone2017,Golovich19a,Golovich19b}. 

Tidal mass loss in clusters can occur through two principal paths. In the first, the cluster mass creates a deep gravitational potential well that can tidally strip satellite galaxies and tidally shock them at pericentre. In the second, galaxies in the cluster suffer repeated high speed flybys, a mechanism known as harassment \citep{Moore1996,Moore1999,Davies2005,Smith2010,Smith2012b,Bialas2015}. The strength of harassment is sensitive to the orbit that galaxies follow \citep{Smith2015}. However, general simulations show that harassment acts as an additional source of mass loss, on top of that produced by the cluster mass, with enhanced mass loss of approximately 20\% in typical cases \citep{Knebe2006,Smith2012,Bialas2015}.

However, not all galaxies fall into clusters in isolation. Some may fall into a group environment prior to reaching the cluster, and thus begin to suffer environmental effects earlier (so called `preprocessing') \citep{Fujita2004,Mihos2004,DeLucia2012,Hou2014,Just2019,Han2018,Jung2018,Sarron2019}. It has been shown by multiple studies \citep{DeLucia2012,Han2018,Donnari2021} that a significant fraction of the cluster population entered the cluster as members of a host during mergers, although the exact fraction is sensitive to the definition of the group members \citep{Berrier2009,Donnari2021}. \cite{Han2018} showed that halos which fall into groups are essentially given an extended period of time in dense environments. As a result, from a sample of strongly tidally stripped halos in clusters, three quarters of them were previously in a host prior to cluster infall. Thus, when a group merges with a cluster, it potentially brings halos that have already suffered tidal mass loss. In fact, \cite{Chun2023} find stars produced by preprocessing in groups contribute significantly to the ICL of unrelaxed clusters. Furthermore, some galaxies may begin to feel the effects of environments directly within the cosmological filaments in which groups tend to be found \citep{Laigle2018}.

This last point is a key issue for the aims of our research. We are interested to understand the impact of major mergers for the survival of subhalos, which correspond to the cluster population. However, it is already well understood that galaxy halos brought in by the merger would influence this because they have been preprocessed in their group. Therefore, instead we aim to study how the complex tidal potentials generated during a major merger enhance the destruction of satellite halos, or subhalos, free from the effects of preprocessing in the group. This is accomplished by measuring the fraction of destroyed subhalos that were inside the virial radius $R_{200}$ of the cluster before the group merger began. In this way, the destruction fractions do not include any halos that may have been preprocessed in the group. As we will demonstrate, this allows us to cleanly quantify the amount of enhanced destruction that occurs due to major mergers of varying mass ratios.

In Sec.~\ref{methods}, we introduce the dark matter only cosmological simulation used for this research, the sample selection criteria, and the measurement of the enhancement in the destruction fraction due to major mergers. In Sec.~\ref{results}, we present and discuss the main results, and we summarize and draw conclusions in Sec.~\ref{conc}.

\begin{table*}
    \centering
    \caption{Properties of the sample clusters of galaxies used in this study. From left to right, the columns are IDs of the host clusters, merger mass ratios $m/M$, lookback time of merger from $z=0$ $T_\mathrm{inf}$, virial mass of the clusters $M_\mathrm{vir} (T_\mathrm{inf})$, virial radius $R_\mathrm{vir} (T_\mathrm{inf})$, concentration $c(T_\mathrm{inf})$, and if the host galaxy of infalling group is destroyed or survives after the collision. The values of the virial mass, virial radius, and concentration are determined at the time of merger $T_\mathrm{inf}$.}
    \label{cl_props}
    \begin{tabular}{lcccccr}
    \hline\hline
    host id                         & $m/M$ & $T_\mathrm{inf}$ [Gyr]& \multicolumn{1}{l}{M$_\mathrm{vir} (T_\mathrm{inf})$ {[}$10^{14}$ M$_\odot${]}} & \multicolumn{1}{l}{R$_\mathrm{vir} (T_\mathrm{inf})$ {[}kpc{]}} & $c(T_\mathrm{inf})$ & \multicolumn{1}{l}{group host} \\ \hline
    50211648                       & 0.2 & 6.0                 & 0.8                                                              & 1015.5                                             & 4.1                 & destroyed                      \\
    50286578                       & 0.4 & 5.9                 & 0.9                                                              & 1036.8                                             & 5.0                 & destroyed                      \\
    50210783                       & 0.3 & 5.7                 & 0.8                                                              & 1018.4                                             & 1.3                 & destroyed                      \\
    50257925                       & 0.3 & 5.6                 & 1.4                                                              & 1209.6                                             & 2.0                 & destroyed                      \\
    50348212                       & 0.4 & 5.3                 & 1.1                                                                & 1122.1                                             & 2.7                 & destroyed                      \\
    50196185                       & 0.4 & 5.2                 & 0.7                                                              & 947.7                                              & 3.8                 & destroyed                      \\
    50273038                       & 0.2 & 4.8                 & 0.8                                                              & 992.8                                              & 5.6                 & destroyed                      \\
    50211225                       & 0.3 & 4.1                 & 0.7                                                              & 926.3                                              & 3.3                 & destroyed                      \\
    50300838 & 0.3 & 3.9                 & 0.9                                                              & 1004.1                                             & 3.2                 & survives                       \\
    \multicolumn{1}{l}{50196812}   & 0.6 & 3.5                 & 0.5                                                              & 808.4                                              & 2.5                 & destroyed                      \\
    50272651                      & 0.3  & 3.5                 & 2.6                                                              & 1428.6                                             & 5.1                 & destroyed                      \\
    50361923                       & 0.6 & 3.3                 & 1.0                                                              & 1036.5                                             & 4.5                 & destroyed                      \\
    \multicolumn{1}{l}{50286853}  & 0.3  & 3.2                 & 1.5                                                              & 1181.6                                             & 5.0                 & destroyed                      \\
    50378733                     & 0.3   & 3.0                 & 3.5                                                              & 1552.4                                             & 2.1                 & survives                       \\
    50165313                       & 0.5 & 2.8                 & 1.4                                                              & 1152.2                                             & 2.2                 & destroyed                      \\
    50438174                      & 0.5  & 2.6                 & 0.8                                                              & 952.7                                              & 2.0                 & survives                       \\
    50348620                      & 0.6  & 2.2                 & 1.0                                                              & 998.0                                              & 3.5                 & destroyed                      \\
    50258408                      & 0.3  & 2.0                 & 0.9                                                              & 954.3                                              & 3.6                 & destroyed                     
    \\ \hline
    \end{tabular}
\end{table*}

\section{Methods}
\label{methods}

\subsection{Cosmological dark matter only simulation}
\label{sec:sim} 
This study is based on a dark matter (DM) only cosmological simulation called the $N-$Cluster simulation. Initial conditions were built using MUSIC \citep{Hahn2013}, and then simulated using $\mathsf{GADGET-3}$ \citep{Springel2005}. Halo finding and tree building were conducted with Rockstar \citep{Behroozi2013} and ConsistentTrees \citep{Behroozi2013}. A number of recent studies have made use of this simulation set to study galaxy evolution, orbits in dense environments, the large scale structure of the Universe and other topics, e.g. \cite{Canducci2022,Chun2022,Chun2023,Jhee2022,Kim2022,Smith2022a,Smith2022b,Awad2023}. For this study, a cubic box of size $120$ Mpc h$^{-1}$ was used, containing 44 galaxy cluster samples, and resolving halos down to a resolution limit of $2 \times 10^{10}\; \mathrm{M}_\odot$. By using the merger tree, we can track the evolution of dark matter halo positions, velocities, masses, virial radii, and other properties and study the merger process. The mass distribution of the sample clusters used in this study ranges from $\sim 1.1 \times 10^{14}\;\mathrm{M}_\odot$ to $4.7 \times 10^{14}\; \mathrm{M}_\odot$ at $z = 0$. We assume the following cosmological parameters $\Omega_M = 0.3$, $\Omega_\Lambda = 0.7$, $H_0 = 68.4$ km s$^{-1}$ Mpc$^{-1}$, $\sigma_8 = 0.816$, and $n = 0.967$.

\subsection{Merging cluster sample selection}
\label{sample_selection}
Amongst the 44 sample clusters obtained from the simulation, a wide range of merger histories can be found, from major mergers to relatively quiet growth histories. To quantify the histories, we measure several quantities for each cluster:

\textit{(i) Time of merger $T_{\mathrm{inf}}$}. The "lookback" time from $z=0$ to when the merger occurred. This is defined as the moment that the centre of the secondary halo crosses the $R_{200}$ of the main cluster, as shown in Fig.~\ref{merger_cartoon}(c). In the case of an infalling group, only the central halo of the group is used to measure $T_{\mathrm{inf}}$.

\textit{(ii) Mass ratio of a major merger $m/M$}. We measure the mass ratio $m/M$ of each merger where $m$ and $M$ refer to the mass of an infalling galaxy, which is the most massive in a group, and that of a host cluster, respectively. For example, a ratio $m/M \geq 1/5$ is where the infalling halo mass is greater than 20\% of the mass of its host cluster at $T_\mathrm{inf}$. This value was chosen to define a major merger. Events with smaller mass ratios were considered minor mergers. Although this choice may appear arbitrary, we tested alternative values and find that mergers that are more minor than this value had only a minor impact on the destruction fraction, as will be shown later in Sec.~\ref{postinfall} and Fig.~\ref{fig:intermediate}.

\textit{(iii) Additional time constraint on when a merger can occur}. We also limited our merger sample to only include merger events occurring between lookback time $6$ and $2$ Gyr prior to $z = 0$. If mergers happen too recently, given the roughly gigayear crossing times of clusters, there is not enough time to see the impact on the cluster population. And at high redshifts, almost all the clusters suffer many complex mergers, therefore we limited the earliest allowed merger. As we will see in Sec.~\ref{postinfall}, there would be little value to considering earlier mergers as the control sample already has such high destruction fractions that it would be difficult to detect an enhancement from a merger.

Based on these criteria for the major mergers, the control sample is identified as follows. Control clusters are not involved in any major mergers from $8$ Gyr to $z = 0$. After examining all the clusters generated by the simulation, $18$ were identified to involve a single major merger and $15$ to form our control sample. Table \ref{cl_props} shows some properties of the selected merging clusters.

\begin{figure*}
    \includegraphics[width=\hsize]{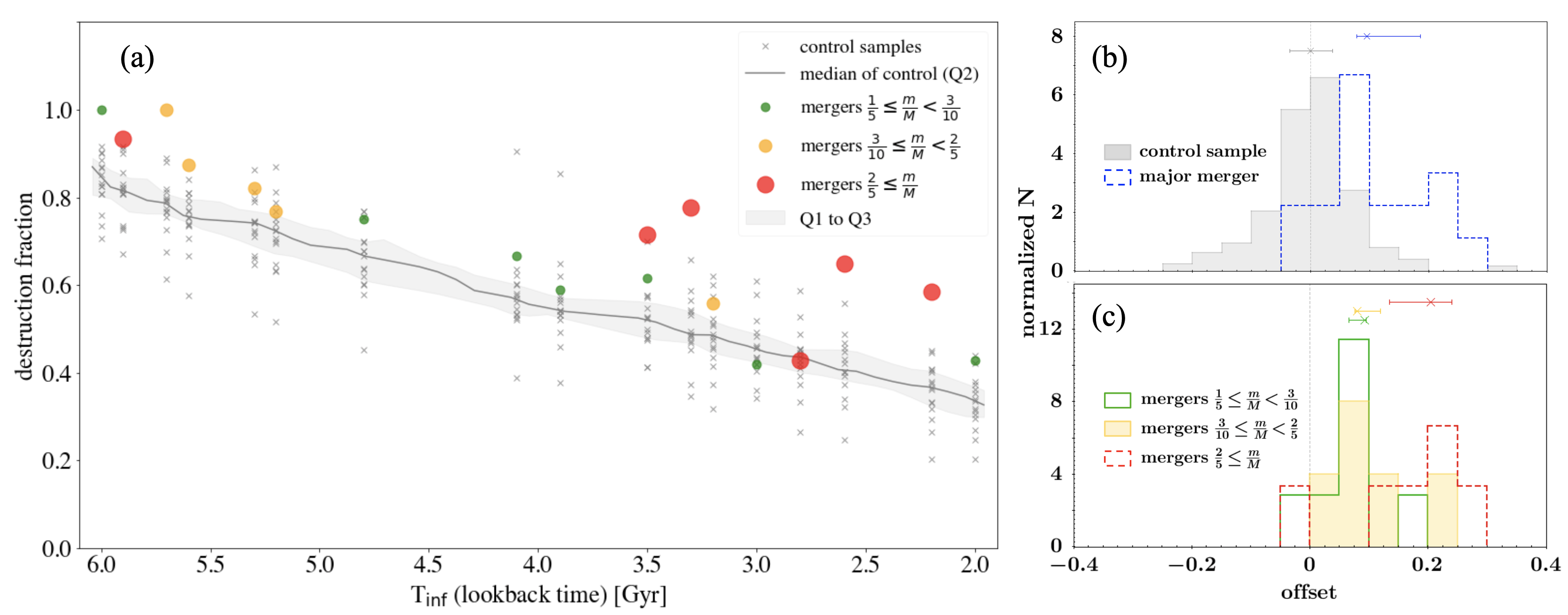}
    \caption{Destruction fraction of clusters plotted as a function of the time of merger $T_\mathrm{inf}$. In Figure 2(a), the green, yellow, and red dots represent the clusters involving major mergers in an order of increasing mass ratios. The grey crosses give the destruction fraction of the control sample at each $T_\mathrm{inf}$ of major mergers for comparison. The grey shaded area indicates the region between the first and third quartiles. The grey line is the median. Figure 2(b) shows the histogram of the offset of the destruction fraction from the median value at a fixed time of merger. The grey filled histogram represents the control sample and the blue dashed one the major mergers. The vertical dashed line is the median value of the control sample and located at zero offset. Error bars show the median and quartiles. Figure 2(c) compares the offset values for different mass ratios of merging clusters with the same color classification used for Figure 2(a). The most major mergers (red dashed histogram) are the most skewed to the right, as the enhancement in destruction is the most extreme.}
    \label{post_infall}
\end{figure*}

\subsection{Destruction of cluster members due to major mergers}

\subsubsection{Cluster member definition}

As noted earlier, we wish to identify cluster members that are inside $R_{200}$ before the merger begins. To do this, we consider all halos inside of 1 $R_{200}$ of the cluster halo. To avoid group members being accidentally included into the cluster member sample, we identify the cluster members 1 Gyr before the merger occurs ($T_\mathrm{inf} + 1$ Gyr in lookback time). We confirm that this criterion successfully removes group members from our cluster member sample. In this way, we are able to study the impact of the merger on the cluster population free from the effects of preprocessing that group members can suffer.

Having identified the cluster members, we can easily track their evolution following the merger down to $z=0$ and measure the destruction fraction over that time period (see following section for details). And we can compare each merger to all of the control sample clusters measured over the same matching time period. This means each merger is compared to all of the control sample clusters, which helps to determine the significance of any enhancement in the destruction fraction.

\subsubsection{Destruction fraction of cluster members}

The destruction fraction is a measure of how many of the cluster members that are identified before the merger ($T_\mathrm{inf} + 1$ Gyr) are destroyed by $z=0$. Since each cluster undergoes merging events at different $T_\mathrm{inf}$, the time intervals between $T_\mathrm{inf} + 1$ Gyr and $z = 0$ differ. However, as we compare to the control sample of clusters and ensure they are compared over exactly the same time period, this is not an issue. The destruction fraction of the control sample is measured in exactly the same way as the merger sample, i.e. measured from an assumed $T_\mathrm{inf} + 1$ Gyr.

\section{Results and discussions}
\label{results}

\subsection{Post-infall destruction fraction of cluster members}
\label{postinfall}

Fig.~\ref{post_infall}(a) shows the destruction fractions for cluster members which undergo major mergers as well as those of the control samples. The x-axis is the time of the merger $T_\mathrm{inf}$ as defined in Sec.~\ref{sample_selection} while the y-axis shows the measure of destruction fraction ranging from $0.0$ to $1.0$. The grey points show the control sample, and the grey line is the median value of the controls while shading indicates the first and third quartile. The median of the destruction fraction clearly decreases with $T_\mathrm{inf}$, because halos are destroyed as they interact in the isolated cluster.

Major mergers are shown by colored points. The mergers are separated into different colour and symbol sizes by how major the merger is, as indicated in the key. It is clear that almost all the mergers lie above the median line, and typically beyond the third quartile of the control sample, meaning the offset is significant. It is apparent that the most major mergers are some of the objects that are lie furthest above the median line.

\begin{figure}
    \centering
    \includegraphics[width=\linewidth]{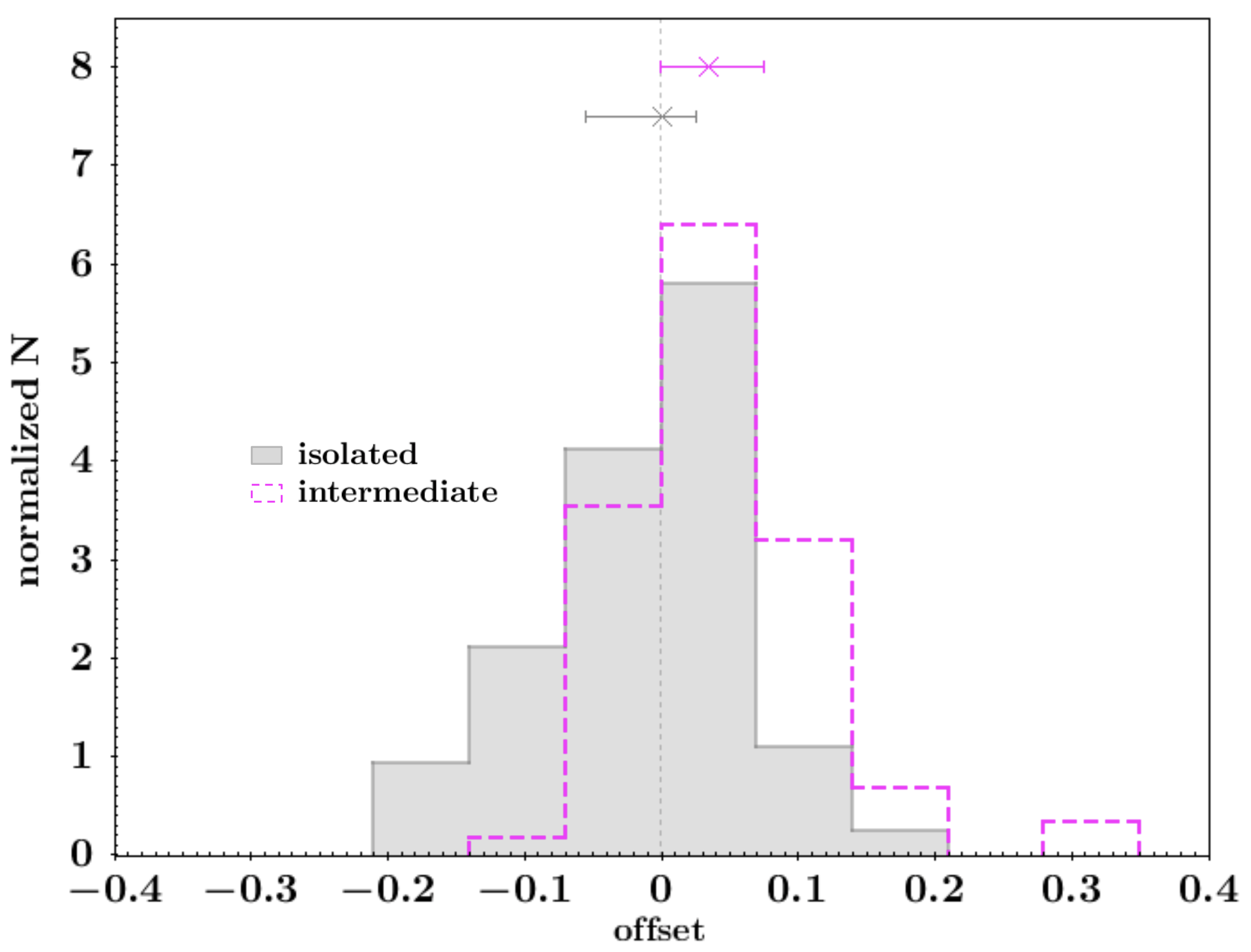}
    \caption{The histograms show the destruction fraction offsets for two different groups of the control sample: isolated (grey filled histograms) and intermediate (magenta dashed histograms) control clusters. The intermediate control clusters are defined to have mergers with mass ratios $m/M$ between $1/7$ and $1/5$ while the isolated control clusters have mass ratios less massive than $1/7$. The isolated control clusters are consistent with no offset. The intermediate subsample shows a small positive offset. Comparing with Fig.~\ref{post_infall}(c), it is clear that the mass ratio must be at least 5:1 to cause significant enhanced destruction, and 5:1 is our chosen limit to be considered a major merger.}
    \label{fig:intermediate}
\end{figure}

To more clearly see the offsets from the median line at a fixed $T_\mathrm{inf}$, we consider the histograms of the offset as shown on the right in Fig.~\ref{post_infall}. In panel (b), the blue dashed line is the histogram of all the merging clusters while the grey filled histogram is the control sample. The cross symbols indicates the median while error bars indicate the quartiles. For the control sample, the median is zero by construction. Meanwhile, the mergers are typically offset to the right by approximately $10\%$, with some objects reaching to higher offsets than  $~30\%$. In order to check if the masses of clusters would influence the destruction measure, we conducted the following test. We first split up the control sample into two subgroups, lower and higher mass control clusters, based on the median value of their masses at $T_\mathrm{inf}$. After reproducing Fig.~\ref{post_infall}(a) and Fig.~\ref{post_infall}(b) by separately plotting the two subgroups, we concluded that there is no significant dependence of cluster masses on the measured destruction fraction offset.

\begin{figure*}
    \includegraphics[width=15cm]{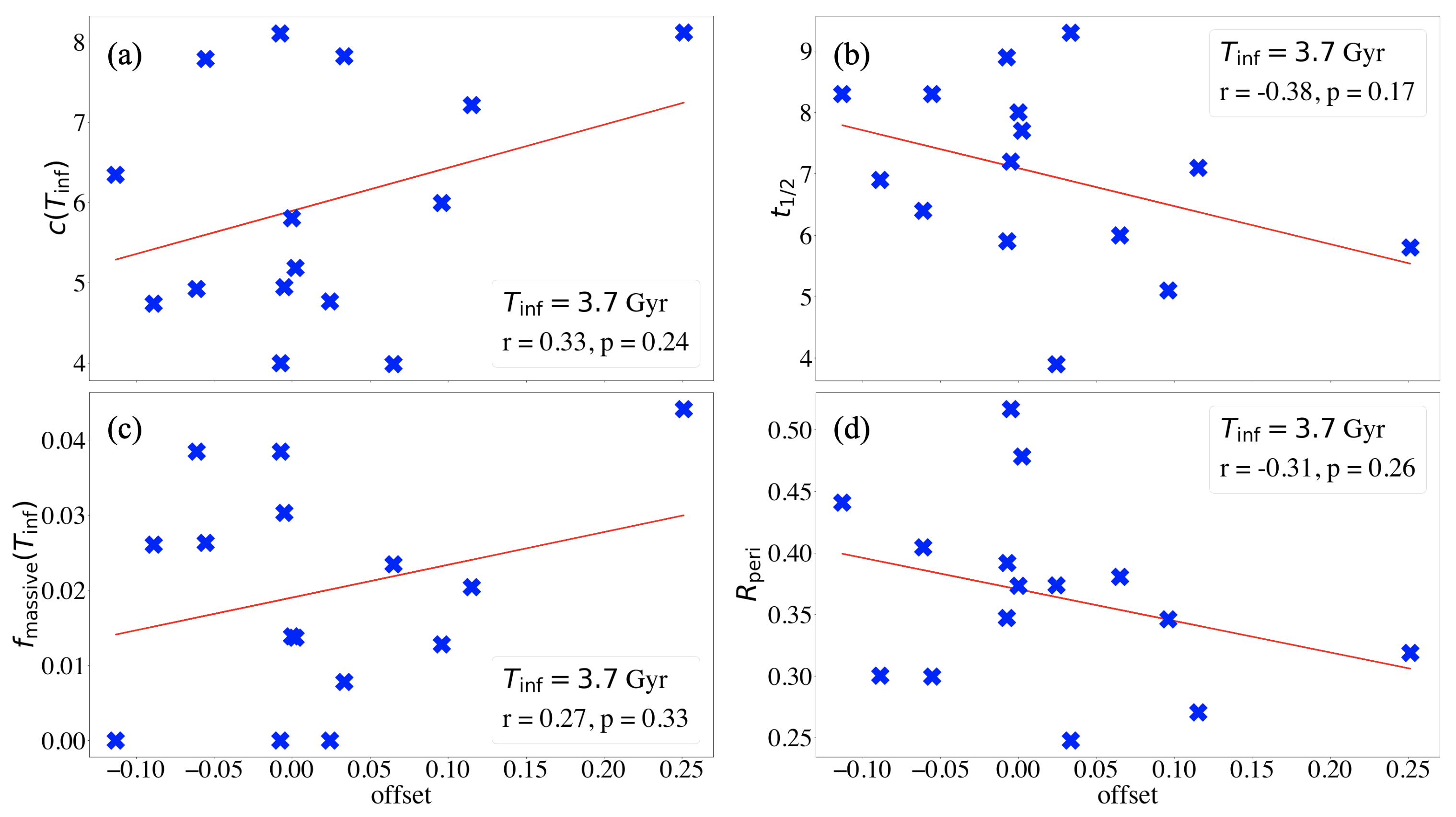}
    \caption{Scatter plots of the correlation between the destruction fraction offsets of the control sample and four different secondary parameters. $T_\mathrm{inf}$ is assumed to be 3.7 Gyr. The secondary parameters are, Figure 4(a) concentration of the cluster measured at the time of merger, 4(b) time since a cluster was half its current day mass, 4(c) fraction of massive halos larger than $3\%$ of their host cluster mass measured at time of merger, and 4(d) the median value of pericentre distance $R_\mathrm{peri}$ for first infaller satellites. 
    The correlation coefficients $r$ and p-values $p$ are measured using least squares regression method and are indicated in each plot. While there is a hint of a correlation by eye, the statistically significance is low with large p-values.}
    \label{secparams}
\end{figure*}

Fig.~\ref{post_infall}(c) shows the offset histogram if we split the mergers in the same way as done in panel (a), and the colours of the histogram match the colours of the points in the legend. The grey vertical line is at $0.0$. The most major mergers (red dashed) are visibly more destructive, with a median offset reaching about 20\%. Even less major mergers, with mass ratios closer to 5:1 (green), can be seen to cause an approximately 10\% enhancement in the destruction fraction of cluster members. This naturally leads to the question, how minor can a merger be in order to cause enhancement?

In order to figure out the critical value for a merger to be called \textit{major}, we tested mergers that were less major than 5:1. Fig.~\ref{fig:intermediate} shows the histograms of offset for the control sample. This time, the control sample was split into two groups: intermediate control and isolated control. Intermediate control clusters, the magenta dashed histogram, are defined to be the mergers with $1/7 \leq m/M < 1/5$ and the isolated control clusters, the grey filled histogram, with $m/M < 1/7$. The isolated sample is consistent with no offset at all when considering the errors. The intermediate sample shows a slight positive offset of similar size as the errors. Comparing with Fig.~\ref{post_infall}(c), we conclude that the critical mass ratio for a merger to cause significant enhanced destruction is 5:1 which is our limiting choice for a major merger.

\subsection{Secondary parameters}

Major mergers have been found to significantly enhance the destruction of cluster members. However, there could be additional factors of significance affecting the results. Thus, we examine four secondary parameters that could potentially play a role in affecting the destruction fraction. The first is the NFW concentration parameter of clusters at the time of merger $c(T_\mathrm{inf})$, where higher concentrations might be expected to increase the destruction fraction. The second is the time a cluster takes to reach its final mass since having half its final mass $t_{1/2}$. Clusters that have rapidly grown their mass might be more likely to produce higher destruction fractions. The third is the fraction of massive substructures inside a cluster which are larger than $3\%$ of the host mass evaluated at the time of merger $f_\mathrm{massive}(T_\mathrm{inf})$. This is a measure of the amount of substructure in the cluster, and in principle higher amounts of substructure could help to tidally disrupt satellites. The fourth and final measure is the median value of the pericentre distance of infalling dark matter halos $R_\mathrm{peri}$. This measurement was made in a previous study \citep{Smith2022a} and we might expect satellites with smaller pericentres to suffer stronger disruption by the cluster tides.

Fig.~\ref{secparams} shows the correlation between these parameters and the destruction fraction offset. 
The best fit for the data given by the correlation coefficient $r$ and the p-value $p$ is computed based on the least squares method. We only consider the control sample for these tests because otherwise the dominating effect of the merger would obscure the impact of the secondary parameter. For consistency, the destruction fractions for the secondary parameters are all computed at a fixed moment of $T_\mathrm{inf} = 3.7$ Gyr, which is the median value of $T_\mathrm{inf}$ for the major merger samples.

From Fig.~\ref{secparams}, it is apparent that all four plots show the correlations close to $\sim 30\%$ with p-values much larger than $0.05$. Thus, we see weak correlations which are statistically insignificant. Still, the results for, for example, $t_{1/2}$ and $R_\mathrm{peri}$ given in Fig.~\ref{secparams}(b) and Fig.~\ref{secparams}(d) do appear to show hints of the expected dependencies described above, and they tend to follow the regression lines, although with multiple outliers. However, the significance of the trends is sensitive to the presence of the point that is furthest on the right in the panels. When we repeat the calculations excluding this point, the weak trends with $c (T_\mathrm{inf})$ and $f_\mathrm{massive}$ disappear with larger p-values of $0.82$ and $0.74$, respectively, and the trends with $R_\mathrm{peri}$ and $t_{1/2}$ are weakened even further with $p = 0.35$ and $p = 0.26$, respectively. We also conducted the same study for $T_\mathrm{inf} = 4.7$ Gyr and $T_\mathrm{inf} = 2.7$ Gyr to see how the correlations change for a different assumption of merger time. The results were found to be on the whole consistent with the one for $T_\mathrm{inf} = 3.7$ Gyr. The plot for $R_\mathrm{peri}$ agrees with the previous study done by \citet{Smith2022a} as smaller $R_\mathrm{peri}$ results in stronger tidal mass loss of host halos. In addition, according to \citet{Wang2020}, the value of concentration fluctuates dramatically {\it during} the merger where the system of the host halo is disturbed by the infalling halo. Indeed, we expect that these dramatic changes in the density field of the host are primarily responsible for the changes to the tidal field that enhance satellite destruction. However, we emphasise that we measure the concentration when the merger is {\it only just beginning} in Fig.~\ref{secparams}(a), and find the concentration is not strongly influential on the satellite destruction fraction at this time.

Lastly, we tried an additional simple test to try to get a better feeling for the relative significance of major mergers compared to the secondary parameters for the destruction fractions. We split the control sample in half with the lower and upper half values of each secondary parameters and then measure how the median offset of the two groups compare. In this way, we get some idea of how much the offset depends on each parameter. For concentration $c(T_\mathrm{inf})$, this gave $\sim0.02$, we also found $\sim0.03$ for $t_{1/2}$, $\sim0.01$ for $f_\mathrm{massive}(T_\mathrm{inf})$, and $\sim0.04$ for $R_\mathrm{peri}$. Thus, $R_\mathrm{peri}$ is the most dominant of the secondary parameters. 

From Fig.~\ref{post_infall}(c), we found that the median values of the destruction fraction offset for the merging clusters are $0.09$, $0.08$, and $0.20$ from the least to the most major mergers. The standard deviation of offset for the control sample $1\sigma \approx 0.09$. Thus, objects from the control sample with an offset of $0.20$ (equal to the median offset of major mergers) are rare, less than two-sigma events. Meanwhile, the computed differences in median for all four secondary parameters are well below $1\sigma$. We note that comparing the upper and lower half of the sample is fairly arbitrary and therefore perhaps should not be directly compared with the offsets from the mergers. Nevertheless, the most major mergers we consider consistently show the largest offsets, where as this would not be true for the secondary parameters according to the scattered trends seen here. And, if other parameters were playing an equally important role as the major mergers, we might not expect to see such a clear offset of the mergers above the control sample median line as shown in Fig.~\ref{post_infall}(a). Therefore, we conclude that the secondary parameters play less significant role on the destruction fraction. 

To summarise, we interpret that the weak correlations do not necessarily rule out the four secondary parameters to affect the destruction of cluster members during major mergers. Instead, they only play a secondary role while the major mergers are found to be the primary factor.

\section{Conclusions}
\label{conc}

To conclude, this study investigated the impact of major mergers on the cluster population, which are inside the virial radius of the cluster before the merger begins, using a DM-only cosmological simulation. To do so, we measure the destruction fraction of the cluster members, the proportion of objects that were cluster members before the merger which are destroyed by $z=0$. It is already well known that galaxy groups can preprocess their satellites before falling into the cluster \citep{Fujita2004,Mihos2004,McGee2009,DeLucia2012,Hou2014,Han2018,Jung2018,Just2019,Sarron2019}. However, here, we are specifically interested to quantify the amount of additional mass loss the members of the cluster suffer due to the destructive tides that arise during the merger process. Therefore, we identify the cluster members prior to the merger (in fact 1 Gyr before the merger begins), and then quantify their destruction fraction by $z=0$. By comparing the measured destruction fractions with a control sample of non-merging clusters, we are able to cleanly measure the additional destruction of cluster members arising from the merger, without the result being influenced by preprocessed halos that are brought in during the merger. In addition, mergers are split by mass ratio in order to see any dependency of destruction fraction on the merger mass ratio. Below is a summary of the main results of this paper.

\begin{itemize}
    \item Major mergers with mass ratios larger than 5:1 cause enhancement above the control sample by approximately 10\% and up to 30\%. The enhancement tends to get bigger if a merger is more major. 
    
    \item We have also checked the minimum merger mass ratio to cause such enhancement. It was found that the mergers with mass ratios smaller than 7:1 did not show any signs of enhancement.

    \item We searched for possible secondary parameters controlling the destruction fraction including halo concentration, time to reach the final mass from half of it, fraction of massive structures, and the median value of pericentre distance for first infallers. Only weak correlations were obtained although they followed the expected trends. In comparison, major mergers are found to be the primary factor for enhancing the destruction of cluster members.
\end{itemize}

These results demonstrate that, in real clusters that undergo major mergers, the strong tidal fields generated during the merger process would be expected to enhance the break down of cluster substructures beyond the enhancement caused by bringing in preprocessed halos into the cluster. Thus, clusters that are undergoing a major merger or have undergone one in the past would be expected to have experienced a period of enhanced tidal destruction of their satellites, resulting in an accelerated growth of their diffuse stellar components in the cluster, including the BCG and the ICL, and potentially presenting an increased fraction of galaxies with disturbed morphologies. However, one limitation of our analysis so far is that we used a DM-only simulation, and therefore we cannot make direct predictions for the response of the stellar components of galaxies to the mergers. However, if a well resolved DM halo is tidally stripped until it is destroyed, it is expected that galaxies will give up their stars into the diffuse ICL, as dark matter and stellar mass losses are intrinsically linked \citep{Smith2016}. 

In any case, an important next step would be to study cluster mergers within the same framework presented here, only using with a larger sample size and based on hydrodynamical cosmological simulations in order to provide more direct predictions for observations, such as the formation and growth of BCGs and ICL \citep{Contreras-Santos2022}. An interesting area of research would be to try to quantify the visibility and frequency of features such as tidal streams caused by the sudden increase in satellite destruction, or barring and tidal distortion of the stellar disks of surviving satellites, as a result of the violent tidal fields generated during major mergers with clusters.

\section*{Acknowledgements}

This work was supported by the National Research Foundation of Korea (NRF) grant funded by the Korea government (MSIT) (2022M3K3A1093827). RS acknowledges support from FONDECYT Regular 2023 \#1230441.

\section*{Data Availability}
The data underlying this article will be shared on reasonable request to the corresponding author.

\bibliographystyle{mnras}
\bibliography{bibfile_Rory}

\bsp
\label{lastpage}
\end{document}